\documentclass[aps,prl,floatfix,tightenlines,superscriptaddress,twocolumn,letterpaper]{revtex4}

\usepackage[bookmarks = true, colorlinks = true, urlcolor=blue]{hyperref}
\usepackage[utf8x]{inputenc}
\usepackage{graphicx}
\usepackage{amsmath}
\usepackage{amsthm}
\usepackage{amsfonts}
\usepackage{amssymb}
\usepackage{paralist}
\usepackage{bbm}
\usepackage{latexsym}
\usepackage{multirow}

\newcommand{\SH}{\mathbf{H}}
\newcommand{\SV}{\mathbf{V}}
\newcommand{\SD}{\mathbf{+}}
\newcommand{\SA}{\mathbf{\!-\!}}
\newcommand{\SL}{\mathbf{L}}
\newcommand{\SR}{\mathbf{R}}
\newcommand{\vac}{\ket{\mathrm{vac}}}
\newcommand{\ket}[1]{|#1\rangle}

\newcommand{\bracket}[2]{\langle #1|#2\rangle}

\begin{document}

\title{Optical one-way quantum computing with a simulated valence-bond solid}
\author{Jonathan Lavoie}
\thanks{These authors contributed equally to this work.}
\affiliation{Institute for Quantum Computing and Department of Physics \&
Astronomy, University of Waterloo, Waterloo N2L 3G1, Canada}
\author{Rainer Kaltenbaek}
\thanks{These authors contributed equally to this work.}
\affiliation{Institute for Quantum Computing and Department of Physics \&
Astronomy, University of Waterloo, Waterloo N2L 3G1, Canada}
\author{Bei Zeng}
\affiliation{Institute for Quantum Computing and the Department of Combinatorics and
Optimization, University of Waterloo, Waterloo N2L 3G1, Canada}
\author{Stephen D. Bartlett}
\affiliation{School of Physics, The University of Sydney,
  Sydney, New South Wales 2006, Australia}
\author{Kevin J. Resch}
\email{kresch@iqc.ca}
\affiliation{Institute for Quantum Computing and Department of Physics \&
Astronomy, University of Waterloo, Waterloo N2L 3G1, Canada}

\begin{abstract}
One-way quantum computation proceeds by sequentially measuring individual spins (qubits) in an entangled many-spin resource state~\cite{Raussendorf2001a}. It remains a challenge, however, to efficiently produce such resource states. Is it possible to reduce the task of generating these states to simply cooling a quantum many-body system to its ground state? Cluster states, the canonical resource for one-way quantum computing, do not naturally occur as ground states of physical systems~\cite{Nielsen2006a, Bartlett2006a}. This led to a significant effort to identify alternative resource states that appear as ground states in spin lattices~\cite{Verstraete2004a,Gross2007a, Gross2007b, Brennen2008a, Chen2009a}. An appealing candidate is a valence-bond-solid state described by Affleck, Kennedy, Lieb, and Tasaki (AKLT)~\cite{Affleck1987a}. It is the unique, gapped ground state for a two-body Hamiltonian on a spin-1 chain, and can be used as a resource for one-way quantum computing~\cite{Verstraete2004a,Gross2007a,Gross2007b,Brennen2008a}. Here, we experimentally generate a photonic AKLT state and use it to implement single-qubit quantum logic gates.
\end{abstract}
\maketitle

In the standard circuit model of quantum computation~\cite{Bennett2000a}, information is carried by two-level systems called qubits. The computation proceeds dynamically via unitary single-qubit logic gates and multiple-qubit entangling gates. Apart from these entangling gates the qubits are fully isolated from each other. Computations in the one-way model, on the other hand, are performed via single-qubit measurements on a strongly-correlated, i.e., entangled, resource state. The one-way model has led to some of the highest estimated error thresholds for fault-tolerant quantum computation~\cite{Raussendorf2006a, Raussendorf2007a}, and to a series of experimental demonstrations of quantum logic gates~\cite{Walther2005a, Prevedel2007a, Vallone2008a, Tokunaga2008a, Biggerstaff2009a, Gao2010a}, wherein the technical requirements can be much simpler than for the circuit model. This is particularly true of optical implementations, where the resource requirements for one-way quantum computing are significantly lower~\cite{Browne2005a}, and the predicted error thresholds significantly higher~\cite{Dawson2006b}, than for any other approach to quantum computation.

\begin{figure}[ht]
  \begin{center}
   \includegraphics[width=0.8\columnwidth]{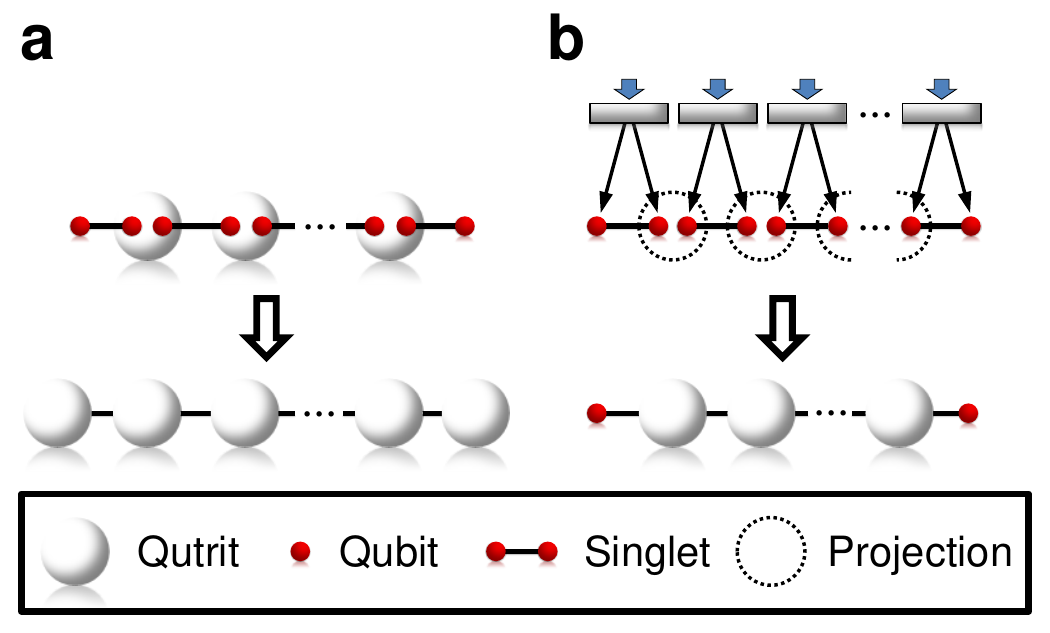}
  \end{center}
 \caption{\textbf{AKLT states.} (a) The AKLT state~\cite{Affleck1987a} is a valence-bond solid and can be represented by a chain of spin-$\frac{1}{2}$ singlet states where adjoining qubits of neighbouring pairs are projected on the triplet subspace, i.e.~the subspace symmetric with respect to swapping of the two qubits. At either end of the chain a boundary qubit remains, ensuring that the ground state is non-degenerate. (b) One can simulate an AKLT state with a chain of sources producing singlet states and projecting pairs of particles on the triplet subspace.\label{scheme}}
\end{figure}

Because qubits in the one-way model are not isolated but rather interact strongly with each other, this approach lends itself more naturally for implementations in condensed-matter systems. But, out of the vast variety of strongly-coupled quantum many-body systems, can we find one that has a ground state we can use as a resource for quantum computing? That seems unlikely if this ground state is to be the cluster state, because the cluster state is \textit{not} the ground state of a strongly-coupled many-body system with a Hamiltonian consisting of two-body interactions~\cite{Nielsen2006a, Bartlett2006a}. As a result, the search for alternative resource states has attracted a lot of interest recently. Although up to now little is known about the requirements potential resource states for the one-way model have to meet, and although most states are in fact useless for this task~\cite{Gross2009a}, a handful of alternative states have been identified~\cite{Verstraete2004a,Gross2007a, Gross2007b, Brennen2008a, Chen2009a}. All of these states can be described in the framework of projected entangled pair states~\cite{Verstraete2004a,Verstraete2008a} or matrix product states~\cite{Gross2007a, Gross2007b}.

\begin{figure}[ht]
  \begin{center}
   \includegraphics[width=\columnwidth]{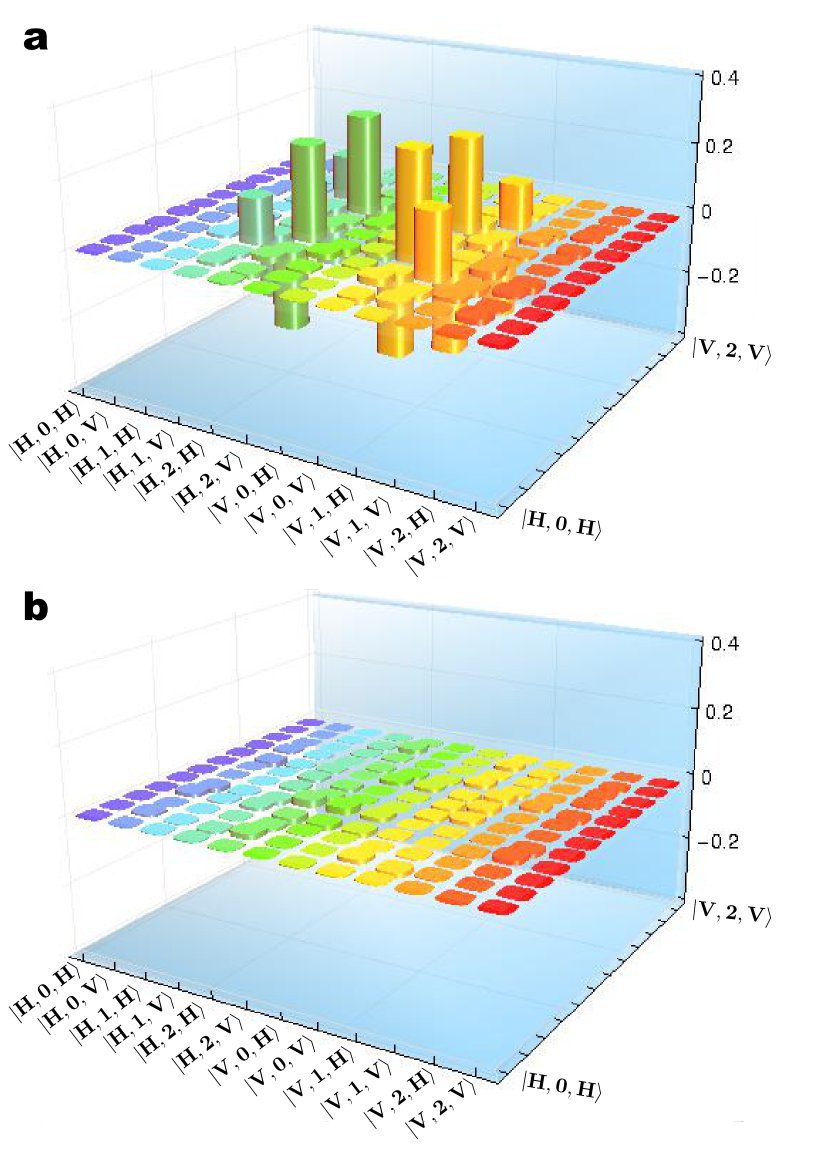}
  \end{center}
 \caption{\textbf{Tomographic reconstruction of our photonic AKLT state.} (a) and (b) show the real and imaginary part of the density matrix reconstructed from an over-complete set of tomography measurements. The fidelity with the ideal AKLT state is $(87.1\pm0.4)$\%.\label{densitymatrix}} 
\end{figure}

A promising candidate is the ground state of a spin model studied by Affleck, Kennedy, Lieb, and Tasaki (AKLT)~\cite{Affleck1987a}. This valence-bond-solid state (see Figure~\ref{scheme}a) appears as the unique gapped ground state of a rotationally-invariant nearest-neighbour two-body Hamiltonian on a spin-1 chain. The AKLT state possesses diverging localisable entanglement length~\cite{Popp2005a} and, remarkably, can serve as a resource for one-way quantum computation~\cite{Gross2007b, Gross2008a, Brennen2008a}. Because the Hamiltonian is frustration free, i.e.~the ground state minimises the energy of each local term of the Hamiltonian, measurements in the course of the computation leave the remaining particles in their ground state. Operations leaving the computational subspace are penalised by the energy gap protecting the AKLT state. Universal quantum computation can be achieved via dynamical coupling of several AKLT states, where each can be regarded as `quantum computational wires'~\cite{Gross2007b, Brennen2008a, Gross2008a}. These properties render the AKLT state an attractive alternative to cluster states as a more natural resource for quantum computing in condensed-matter systems.

Quantum computation with AKLT states is different from computing with cluster states in a number of ways. The elementary physical units are spin-1 systems (qutrits) instead of spin-$\frac{1}{2}$ systems (qubits), although it is still qubits that are encoded as `logical' information. Adaptive measurements allow the performance of non-Pauli operations, including Clifford gates. Single-qubit rotations can be performed around any Cartesian axis. These operations are probabilistic, rather than deterministic, and succeed with probability $\frac{2}{3}$. When an operation fails, it performs a heralded logical-identity operation, i.e.~a teleportation of the logical information along the chain. The operation can then be reattempted on the next qutrit until it succeeds. Combinations of such rotations allow the implementation of arbitrary single-qubit quantum logic gates.

Although a number of one-dimensional spin chains are well-described by the AKLT Hamiltonian, most prominently Ni(C$_2$H$_8$N$_2$)$_2$NO$_2$(ClO$_4$) (NENP)~\cite{Hagiwara1990a}, up to now experimental techniques do not allow the single-spin measurements necessary for one-way quantum computation. Yet, one of the fundamental and most appealing motivations for quantum computing, is the possibility to simulate aspects of quantum  systems that cannot directly be studied~\cite{Buluta2009a}. Because the AKLT state is a valence-bond solid state (see Figure~\ref{scheme}a), we can simulate it via a chain of spin-$\frac{1}{2}$ singlet states, for example polarisation-entangled photon pairs, where adjoining particles of neighbouring pairs are projected on the symmetric triplet subspace (see Figure~\ref{scheme}b). While this approach does not allow to analyse the dynamics of the corresponding solid-state system, it allows the direct production of the AKLT state, and to use it for one-way quantum computation.

\begin{figure}[ht]
  \begin{center}
   \includegraphics[width=\columnwidth]{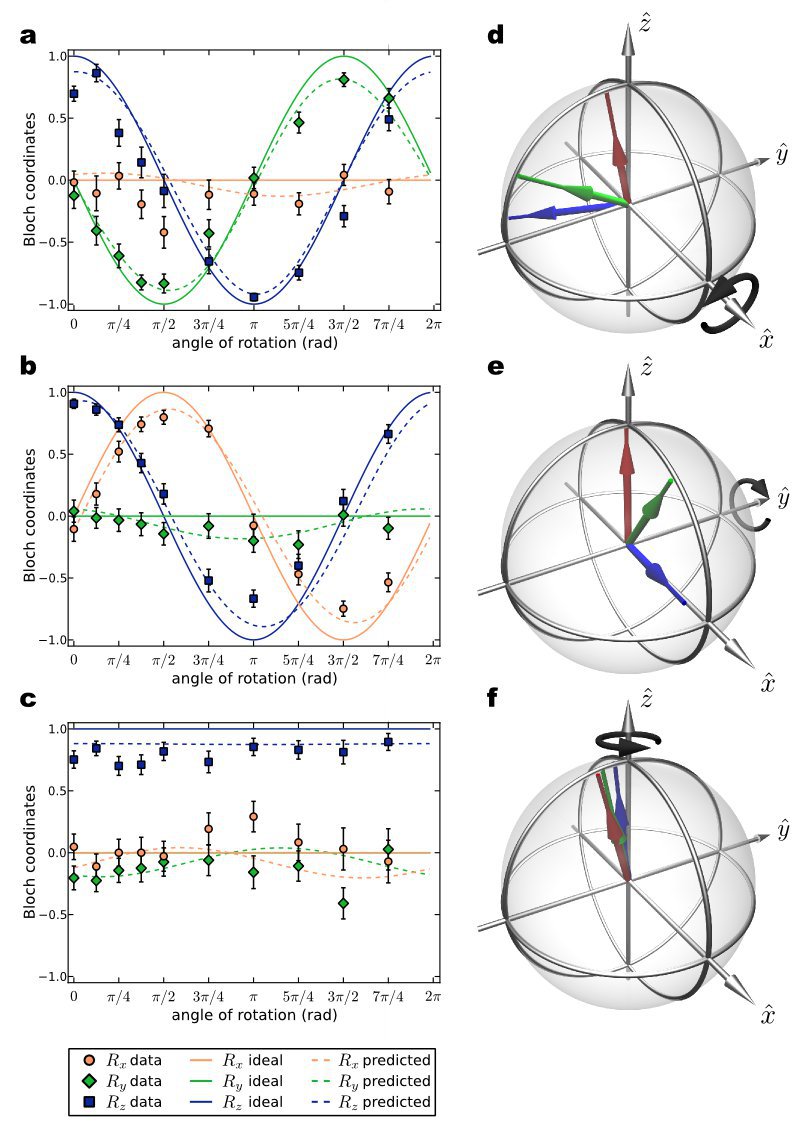}
  \end{center}
 \caption{\textbf{Measurement results for single-qubit rotations.}. (a)-(c) show the coordinates of the Bloch vectors of the reconstructed output density matrices for rotations of a logical input state $\ket{\SH}$ around the $\hat{x}$, $\hat{y}$ and $\hat{z}$ axes, respectively. Note that the results shown are for the `plus' outcome of the qutrit measurement, and that we have applied the necessary Pauli corrections to the reconstructed density matrices for all plots shown in the figure. Error bars are standard deviations calculated from Monte-Carlo simulations. Solid and dashed lines indicate the theoretical expectations given the ideal AKLT state and the tomographically-reconstruced AKLT state (see Figure~\ref{densitymatrix}), respectively. For the rotation angles $0$, $\frac{\pi}{4}$, and $\frac{\pi}{2}$, panels (d)-(f) show the Bloch vectors of the measured (and Pauli corrected) density matrices corresponding to the Bloch coordinates shown in (a)-(c).\label{rotfig}}
\end{figure}

Here, we experimentally demonstrate the generation of photonic AKLT states and their application for one-way quantum computation. We produce two singlet states, $\ket{\psi^-} = (\ket{\SH\SV}-\ket{\SV\SH})/\sqrt{2}$, in four distinct spatial modes using spontaneous parametric down conversion. Here, $\ket{\SH}$ and $\ket{\SV}$ denote horizontal and vertical polarisation, our computational basis. From these two singlets we create an AKLT state consisting of two boundary spin-1/2 systems and one spin-1 system. The spin-1 system is a biphoton, symmetrised by projecting a pair of photons into the triplet subspace. Qutrit measurements are performed using the method from Ref.~\cite{Lanyon2008a}. For details of our experimental setup, see the Methods section and Appendix \ref{app:exp}. A detailed discussion of the theoretical aspects of the simulation of AKLT states using quantum optics and their use in one-way quantum computation will be given in a separate paper~\cite{Darmawan2010a}.

The AKLT state for a qubit-qutrit-qubit system is $\ket{\psi_{AKLT}} = \frac{1}{\sqrt{6}}\ket{\SH,\mathbf{1},\SV} + \frac{1}{\sqrt{6}}\ket{\SV,\mathbf{1}, \SH}-\frac{1}{\sqrt{3}}\ket{\SH,\mathbf{2},\SH}-\frac{1}{\sqrt{3}}\ket{\SV,\mathbf{0},\SV}$. In our case, the qutrit states $\ket{\mathbf{0}}$, $\ket{\mathbf{1}}$, $\ket{\mathbf{2}}$ correspond to the biphoton states $\frac{1}{\sqrt{2}} a^\dagger_\SH a^\dagger_\SH \vac$, $a^\dagger_\SH a^\dagger_\SV \vac$, $\frac{1}{\sqrt{2}} a^\dagger_\SV a^\dagger_\SV \vac$, respectively. Here, $\vac$ is the vacuum state, and $a^\dagger_{\SH}$ and $a^\dagger_{\SV}$ are photon creation operators. To verify the faithful production of the AKLT state in our experiment, we perform quantum-state tomography on the qubit-qutrit-qubit system and use a maximum-likelihood technique based on a semi-definite-programming algorithm \cite{Doherty2009a} to reconstruct the density matrix shown in Figure~\ref{densitymatrix}. For a detailed list of the states measured and of the corresponding counts, see the Table \ref{tomocounts} in Appendix \ref{app:results}. The fidelity of the reconstructed density matrix with the ideal AKLT state is $(87.1\pm0.4)$\%. We calculate this value in a Monte-Carlo simulation with $420$ iterations on the observed counts, using the definition $F(\rho,\sigma)=\left(\mathrm{Tr}\sqrt{\sqrt{\sigma} \rho \sqrt{\sigma}}\right)^2$ for the fidelity between two quantum states~\cite{Jozsa1994a}.

\begin{table*}[ht]
 \begin{center}
  \begin{tabular}{cccccccccc}
    \hline\hline
     \multirow{2}{*}{Rot.}& \hspace{0.3em} & \multicolumn{2}{c}{plus}&\hspace{0.3em} & \multicolumn{2}{c}{minus}&\hspace{0.3em} & \multicolumn{2}{c}{id}\\[0.3ex]\cline{3-4}\cline{6-7}\cline{9-10}
    & & state & corr.& & state & corr.& & state & corr. \\[0.3ex]\hline
    $R_x(\theta)$ & & $\cos\frac{\theta}{2}\ket{y}+i \sin\frac{\theta}{2}\ket{z}$ & $Y$ & & $i\sin\frac{\theta}{2}\ket{y} + \cos\frac{\theta}{2}\ket{z}$ & $Z$ & & $\ket{x}$ & $X$ \\[0.3ex]
    $R_y(\theta)$ & & $\cos\frac{\theta}{2}\ket{z} + \sin\frac{\theta}{2}\ket{x}$ & $Z$ & & $-\sin\frac{\theta}{2}\ket{z} + \cos\frac{\theta}{2}\ket{x}$ & $X$ & & $\ket{y}$ & $Y$ \\[0.3ex]
    $R_z(\theta)$ & & $\cos\frac{\theta}{2}\ket{x}+i \sin\frac{\theta}{2}\ket{y}$ & $X$ & & $i\sin\frac{\theta}{2}\ket{x} + \cos\frac{\theta}{2}\ket{y}$ & $Y$ & & $\ket{z}$ & $Z$ \\[0.3ex]\hline\hline
  \end{tabular}
  \caption{\textbf{Qutrit measurement bases and Pauli corrections.} Single-qubit rotations are realised by a projective measurement in a corresponding qutrit basis that has three possible outcomes: `plus', `minus' and `id'. The qutrit states $\ket{x}$, $\ket{y}$, and $\ket{z}$ are defined as $\frac{1}{2}\left(a^\dagger_\SH a^\dagger_\SH - a^\dagger_\SV a^\dagger_\SV\right)\vac$, $\frac{1}{2}\left(a^\dagger_\SH a^\dagger_\SH + a^\dagger_\SV a^\dagger_\SV\right)\vac$, and $\frac{1}{\sqrt{2}}\left(a^\dagger_\SH a^\dagger_\SV + a^\dagger_\SV a^\dagger_\SH\right)\vac$, respectively, and $X$, $Y$, $Z$ indicate the Pauli correction that has to be applied to the read-out qubit depending on measurement outcome and measurement basis. \label{rotbases}}
 \end{center}
\end{table*}

In order to demonstrate the use of AKLT states for quantum computation, we realise single-qubit rotations around the $\hat{x}$, $\hat{y}$ and $\hat{z}$ axis of the Bloch sphere. We begin the computation by projecting the first boundary qubit onto some qubit state $\ket{\psi}$. By doing so, we effectively prepare the logical state $\ket{\psi^\bot}$, where $\bracket{\psi}{\psi^\bot}=0$. To perform a rotation $R_x(\theta)$, $R_y(\theta)$ or $R_z(\theta)$ of the logical state by an angle $\theta$ around the respective coordinate axis, we project the qutrit on one of the corresponding bases given in Table~\ref{rotbases}. Qutrit measurements are realised probabilistically as described in the Method section and Appendix \ref{app:exp}.  We denote the three outcomes of each qutrit basis as `plus', `minus' and `id'. Each is expected to occur with probability $1/3$. Up to a known Pauli error~\cite{Raussendorf2001a,Gross2007b,Brennen2008a}, which can be corrected as indicated in Table~\ref{rotbases}, the outcomes `plus' and `minus' signal a successful rotation, and the outcome `id' signals the logical identity, i.e., a rotation by $0^\circ$. As a result, a successful rotation is achieved with probability $2/3$. For $\theta = 0$, every outcome heralds the logical identity. This can be used to teleport logical information along the wire, for example to a position where the wire is coupled to another, or to the read-out position.

To prepare our logical input state, we project the first qubit on one of a set of states: {$\ket{\SH}$, $\ket{\SV}$, $\ket{\mathbf{\pm}}$, $\ket{\mathfrak{h}^\pm}$, $\ket{\mathfrak{m}^\pm}$}. Here, $\ket{\mathbf{\pm}}=(\ket{\SH}\pm\ket{\SV})/\sqrt{2}$, $\ket{\mathfrak{h}^\pm}$ are the eigenstates of the Hadamard operator, $\ket{\mathfrak{m}^+}=\cos(\frac{\xi}{2})\ket{\SH}+\mathrm{e}^{i \frac{\pi}{4}} \sin(\frac{\xi}{2})\ket{\SV}$ is the `magic state' \cite{Raussendorf2007a} with $\xi = \mathrm{arccos}(\frac{1}{\sqrt{3}})$, and $\bracket{\mathfrak{m}^-}{\mathfrak{m}^+}=0$. For each axis of rotation we choose $10$ angles $\theta = \left\{0, \frac{\pi}{8}, \frac{\pi}{4}, \frac{3 \pi}{8}, \frac{\pi}{2}, \frac{3\pi}{4}, \pi, \frac{5\pi}{4}, \frac{3\pi}{2}, \frac{7\pi}{4} \right\}$ and project the qutrit on the corresponding state (see Table~\ref{rotbases}) for the `plus' and `minus' outcomes. We project on the `id' outcome once for every input state and rotation axis. Finally, we reconstruct the density matrix of the computational outcome by performing a tomographically over-complete set of measurements on the last qubit, using the measurement settings: $\ket{\SH}$, $\ket{\SV}$, $\ket{\mathbf{\pm}}$, $\ket{\SR}=(\ket{\SH}+i \ket{\SV})/\sqrt{2}$, and $\ket{\SL}=(\ket{\SH} - i \ket{\SV})/\sqrt{2}$.

Figure~\ref{rotfig} shows measurement results for single-qubit rotations of the logical input state $\ket{\SH}$ (i.e.~projecting the first qubit on $\ket{\SV}$) around the three rotation axes. The plots in Figure~\ref{rotfig}a-c) show the coordinates of the rotated Bloch vectors as compared with the theoretical expectation. In Table~\ref{fidTable} we list the fidelities for rotations of $\ket{\SH}$ as well as the averaged fidelities for all logical input states prepared. The probabilities for the three qutrit measurements, averaged over all input states, rotations and rotation angles, are measured to be $0.34\pm0.03$, $0.30\pm0.05$, and $0.36\pm0.04$ for the `plus', the `minus', and the `id' outcome, respectively. This is in good agreement with the expected value of $\frac{1}{3}$ for each outcome. An average of the output fidelities achieved over all input states and all rotations performed yields a value of $(92\pm4)$\%, demonstrating the high quality of our single-qubit quantum logic gates using a photonic AKLT state. A detailed list of all results can be found in Appendix \ref{app:results}.

\begingroup
\begin{table*}[ht]
 \begin{center}
  \small{\begin{tabular}{cccccccccc}
    \hline\hline
    \multicolumn{10}{c}{gate fidelities for logical input $\ket{\SH}$} \\[0.3ex]\hline
    \multirow{2}{*}{outcomes}& \hspace{0.5em} & \multicolumn{2}{c}{$R_x$} & \hspace{0.3em} & \multicolumn{2}{c}{$R_y$} & \hspace{0.3em} & \multicolumn{2}{c}{$R_z$} \\[0.3ex]\cline{3-4}\cline{6-7}\cline{9-10}
    & & $\rho_{th}$ & $\rho_{exp}$ & & $\rho_{th}$ & $\rho_{exp}$ & & $\rho_{th}$ & $\rho_{exp}$ \\[0.3ex]\hline
    plus & & $0.91\pm0.04$ & $0.98\pm0.02$ & & $0.90\pm0.05$ & $0.98\pm0.02$ & & $0.90\pm0.03$ & $0.98\pm0.02$ \\[0.3ex]
    minus & & $0.93\pm0.03$ & $0.97\pm0.03$ & & $0.91\pm0.03$ & $0.99\pm0.01$ & & $0.92\pm0.04$ & $0.97\pm0.02$ \\[0.3ex]
    id & & $0.90\pm0.03$ & $0.98\pm0.02$ & & $0.92\pm0.02$ & $0.999\pm0.006$ & & $0.97\pm0.02$ & $0.99\pm0.02$ \\[0.3ex]\hline\hline
    \multicolumn{10}{c}{gate fidelities averaged over all input states} \\[0.3ex]\hline
    \multirow{2}{*}{outcomes}& \hspace{0.5em} & \multicolumn{2}{c}{$R_x$} & & \multicolumn{2}{c}{$R_y$} & & \multicolumn{2}{c}{$R_z$} \\[0.3ex]\cline{3-4}\cline{6-7}\cline{9-10}
    & & $\rho_{th}$ & $\rho_{exp}$ & & $\rho_{th}$ & $\rho_{exp}$ & & $\rho_{th}$ & $\rho_{exp}$ \\[0.3ex]\hline
    all & & $0.92\pm0.04$ & $0.97\pm0.02$ & & $0.91\pm0.04$ & $0.98\pm0.01$ & & $0.92\pm0.04$ & $0.98\pm0.02$ \\[0.3ex]\hline\hline
   \end{tabular}}
  \caption{\textbf{Single-qubit logic gate fidelities.} We compare the experimentally determined output density matrices with the ones expected, on the one hand, given an ideal AKLT state, $\rho_{th}$, and, on the other hand, given the AKLT state measured in our setup, $\rho_{exp}$. The upper part of the table shows the fidelities for a logical input state $\ket{\SH}$. For the `plus' and `minus' outcomes the fidelities are averaged over all rotation angles, for the `id' outcome we performed one measurement per rotation axis. The lower part shows the corresponding fidelities averaged over all logical input states prepared (see text) and over all three qutrit measurement outcomes.\label{fidTable}}
 \end{center}
\end{table*}
\endgroup

We have experimentally demonstrated a one-way quantum-computation scheme harnessing a novel resource, the AKLT state, and used it to implement a circuit realising single-qubit rotations around any coordinate axis. Quantum computation using AKLT instead of cluster states promises to combine the inherent advantages of the one-way model with resources that occur naturally in physical systems. Our scheme for creating AKLT states uses entangled states and linear optics similar in requirements to optical implementations using cluster states~\cite{Browne2005a}. In contrast to some other optical implementations of quantum logic gates for one-way quantum computation~\cite{Walther2005a, Prevedel2007a}, our scheme does not require phase stability and achieves significantly higher experimental fidelities. Our implementation of a valence-bond-solid state is a realisation of a projected entangled pair state~\cite{Verstraete2008a}. Such states offer a promising framework for understanding the properties of entangled states that make them useful computational resources~\cite{Verstraete2004a,Gross2007a,Gross2007b,Chen2009a}. Generalisations of the presented approach might allow to simulate other classes of alternative resource states with linear optics and to study their potential for quantum computing. Future challenges will be to find efficient methods of coupling quantum wires, to study solid-state compounds with ground states that can be used as computational resources, and to implement techniques to address such systems on a single-particle level. Ideally, this and related research will lead to implementations in solid-state architectures, allowing to tap the power of one-way quantum computation while taking full advantage of the appealing characteristics of novel resource states like AKLT.

We thank W. A. Coish, R. Prevedel, A. C. Doherty and A. Gilchrist for valuable discussions. We are grateful for financial support from NSERC, OCE, CFI, QuantumWorks, and MRI ERA.

\section*{Methods}
The light source in our experiment is a Titanium:Sapphire femtosecond laser, centred at $790$~nm with $10$~nm full-width-at-half-maximum (FWHM) bandwidth, $2.9$~W average output power and $80$~MHz repetition rate. Second-harmonic generation in a $2$~mm thick Bismuth-Borate crystal (BiBO) yields a beam of $780$~mW power, centred at $395$~nm, with about $1.5$~nm FWHM bandwidth. With this beam we pump two separate type-I spontaneous parametric down-conversion sources \cite{Kwiat1999a, Lavoie2009a}, each a pair of $1$~mm thick beta-Barium-Borate (BBO) crystals. Longitudinal and transverse walk-off occurring in the down-conversion crystals is compensated with a combination of birefringent crystals ($\alpha$-BBO, quartz and BiBO, see Ref.~\cite{Lavoie2009a} and Appendix \ref{app:exp}). All photons pass through $3$nm FWHM bandwidth filters. In each source, the polarisation of the photons in one mode is measured directly at the source, the photons in the other modes are coupled into single-mode fibres and sent to a quantum interferometer and analyser setup. The input modes of the interferometer are overlapped at a 50:50 beam splitter (BS), where, depending on the two-photon state, two-photon interference leads to both photons leaving via the same or via different BS output modes~\cite{Mattle1996a}. By measuring a two-photon event in one output mode of the BS, the biphoton is projected onto a qutrit subspace. This mode is input in a qutrit analyser, where we implement qutrit projections by probabilistically separating the two photons at another BS and performing appropriate polarisation measurements on each photon~\cite{Lanyon2008a}. If we assume that the two polarisation measurements project on the states $\ket{\psi_m} = \left(\cos\alpha_m a^\dagger_\SH+e^{i\chi_m}\sin\alpha_m a^\dagger_\SV\right)\vac$ ($m=1,2$), we can calculate the qutrit state this measurement projects on by propagating these states back through the BS. For a more detailed discussion of the setup and the qutrit projections, see Appendix \ref{app:exp}.

\appendix

\section{Experimental Setup}
\label{app:exp}

\begin{figure*}
 \begin{center}
  \includegraphics[width=0.8\linewidth]{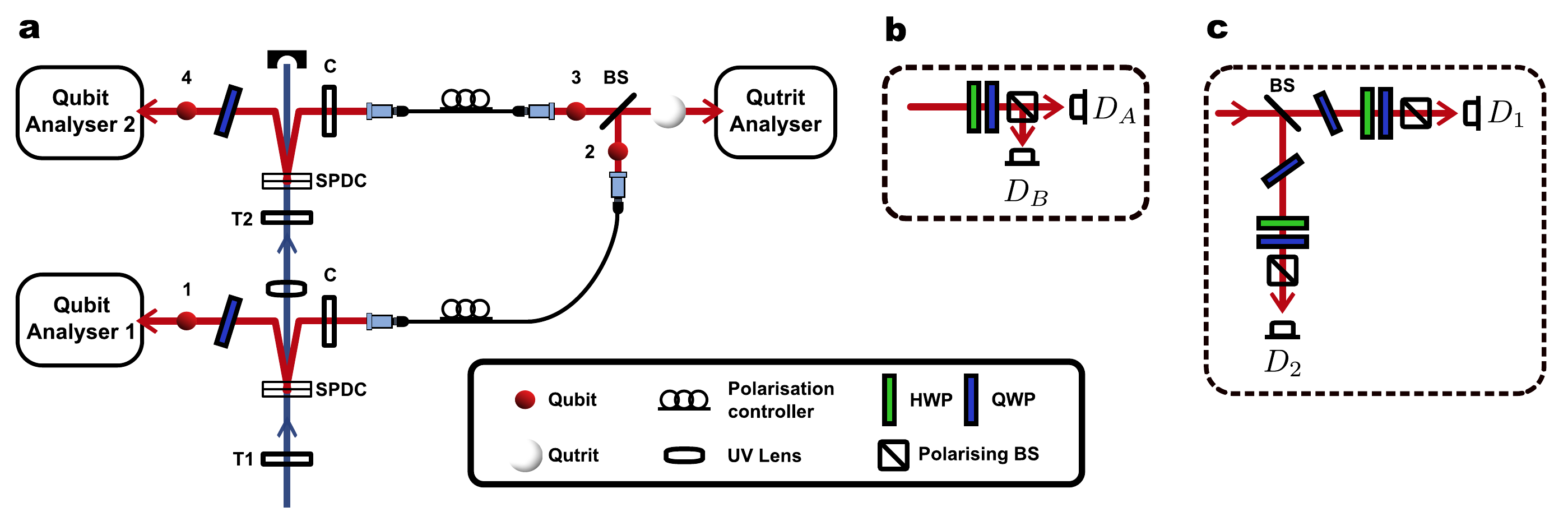}\\
 \end{center}
 \caption{\textbf{Experimental Setup.} (a) Two type-I SPDC sources are used to generate entangled pairs. Longitudinal and transverse walk-off are compensated via combinations of birefringent crystals T1, T2 and C. All photons are coupled into single-mode fibers. Polarisation rotation in the fibres is compensated via polarisation controllers, and the phase is adjusted by tilting quarter waveplates (QWP). Photons in modes $2$ and $3$ are fed into a 50:50 beam splitter (BS). Detecting two photons in one of the BS outputs projects these photons on the symmetric qutrit subspace. They can then be treated as a qutrit. (b) Each qubit analyser consists of a half waveplate (HWP), a QWP and a polarising BS. Both outputs of the polarising BS are monitored. (c) In the qutrit analyser the biphoton forming the qutrit is probabilistically split up using a BS. Phases introduced at the BS are compensated by tilted QWPs. A combination of QWPs and HWPs allows to project the qutrit onto a state of our choice. A successful projection is heralded by a coincidence event between detectors $D_1$ and $D_2$.\label{setup}}
\end{figure*}

The entangled photon pairs in our experiment are generated using two separate type-I spontaneous parametric down conversion (SPDC) sources \cite{Kwiat1999a, Lavoie2009a}. Each consists of a pair of $1$~mm thick beta-Barium-Borate ($\beta$-BaB$_2$O$_4$, or BBO) crystals, their optical axes oriented perpendicular to each other. Longitudinal walk off in the SPDC crystals is compensated using a $0.5$mm quartz, a $2$mm quartz and a $1$mm $\alpha$-BBO crystal before the first SPDC source, and a $2$mm $\alpha$-BBO and a $1$mm quartz crystal between the two sources. Additional transverse walk-off is compensated by placing $1$mm thick BiB$_3$O$_6$ crystals cut at $\theta = 152.6^\circ$ and $\phi=0^\circ$ in modes $2$ and $3$ (see Figure~\ref{setup}a). These angles are chosen such that the crystals compensate transverse walk-off without introducing longitudinal walk off. All photons pass through $3$nm FWHM bandwidth filters. The phases in the setup and the polarisation rotation in the single-mode fibres is set such that the sources produce singlet states in modes 1 \& 2 and 3 \& 4. In modes 1 \& 2 we measure a fidelity of $(96.9\pm0.5)$\% with the ideal singlet state $\ket{\psi^-} = \frac{1}{\sqrt{2}}(\ket{\SH \SV}-\ket{\SV \SH})$, and a tangle of $0.92\pm 0.01$. For the second source the fidelity is $(96.9\pm0.6)$\% and the tangle is $0.90\pm 0.01$. We had single count rates of about $200$kHz in the qubit analysers $1$ and $2$ (see Figure~\ref{setup}a), and single count rates of around $80$kHz in the detectors $D_1$ and $D_2$ in the qutrit analyser (see Figure~\ref{setup}c) (both sources contribute to these latter single count rates). The two-fold coincidence count rate for the first source was $7.4$kHz between qubit analyser $1$ and $D_1$ in the qutrit analyser. For the second source we had a two-fold coincidence count rate of $5.9$kHz between qubit analyser $2$ and $D_1$ in the qutrit analyser.

One photon of each pair is measured directly at the source, using polarisation analysers. The modes for both measurement outcomes are coupled into single-mode fibres and monitored via single-photon detectors (Perkin-Elmer, SPCM-4Q4C). The two remaining photons are coupled into single-mode fibres and sent to a quantum interferometer and analyser setup. The input modes of the interferometer are overlapped at a 50:50 beam splitter (BS). If the input photons are set to have the same polarisation, Hong-Ou-Mandel (HOM) interference \cite{Hong1987a} occurs. Postselecting on four-fold events with one photon in mode $1$, one in mode $4$ and two photons in the output mode of the BS indicated in Figure~\ref{setup}a, we observe constructive HOM interference with a visibility of $95.7\pm 3.7$\%.

\begin{table*}
 \begin{center}
 \begin{tabular}{cccccccccccccccc}
    \hline\hline
    \multirow{2}{*}{Rot.} & \hspace{0.5em} & \multicolumn{4}{c}{plus} & \hspace{0.7em} & \multicolumn{4}{c}{minus} & \hspace{0.7em} & \multicolumn{4}{c}{id} \\\cline{3-6}\cline{8-11}\cline{13-16}
    & & $\alpha_1$ & $\chi_1$ & $\alpha_2$ & $\chi_2$ & & $\alpha_1$ & $\chi_1$ & $\alpha_2$ & $\chi_2$ & & $\alpha_1$ & $\chi_1$ 
		& $\alpha_2$ & $\chi_2$ \\[0.3ex]\hline
    $R_x(\theta)$ & &
    $\frac{\theta - \pi}{4}$ & $\frac{\pi}{2}$ & $\frac{3 \pi-\theta}{4}$ & $-\frac{\pi}{2}$ &  &
    $\frac{\theta}{4}$ & $\frac{\pi}{2}$ & $\frac{\pi}{2}-\frac{\theta}{4}$ & $-\frac{\pi}{2}$ &  &
    $\frac{\pi}{4}$ & $\pi$ & $\frac{3 \pi}{4}$ & $-\pi$ \\[0.3ex]
    $R_y(\theta)$ &  &
    $\frac{\theta}{4}$ & $\pi$ & $\frac{\pi}{2}+\frac{\theta}{4}$ & $-\pi$ &  &
    $\frac{\pi-\theta}{4}$ & $0$ & $\frac{3 \pi-\theta}{4}$ & $0$ &  &
    $\frac{\pi}{4}$ & $\frac{\pi}{2}$ & $\frac{\pi}{4}$ & $-\frac{\pi}{2}$ \\[0.3ex]
    $R_x(\theta)$ & &
    $\frac{-\pi}{4}$ & $-\frac{\theta}{2}$ & $\frac{\pi}{4}$ & $-\frac{\theta}{2}$ &  &
    $\frac{-\pi}{4}$ & $\frac{\pi-\theta}{2}$ & $\frac{\pi}{4}$ & $\frac{\pi-\theta}{2}$ &  &
    $0$ & $\frac{\pi}{2}$ & $\frac{\pi}{2}$ & $-\frac{\pi}{2}$ \\[0.3ex]\hline\hline
 \end{tabular}
 \caption{Analyser parameters for qutrit measurements for rotation gates. In the qutrit analyser, one photon is projected on $\ket{\psi_1}$, the second one on $\ket{\psi_2}$, where $\ket{\psi_m} = \cos\alpha_m \ket{\SH}+e^{i\chi_m}\sin\alpha_m\ket{\SV}$. Each qutrit measurement has three possible outcomes, `plus', `minus' and `id' corresponding to three different sets of parameters for the qutrit analyser. We provide the settings for rotations around each of the Cartesian axes.\label{Q3RotSettings}}
 \end{center}
\end{table*}

Whether two photons entering the interferometer (via modes $2$ and $3$) leave through the same or different BS outputs, depends on the biphoton state \cite{Mattle1996a}. In particular, coincidence detection events between two different BS outputs only occur for the two-photon singlet state. By post-selecting on a biphoton excitation in one output mode of the BS, the biphoton is projected onto a symmetric subspace and can be described as a qutrit~\cite{Lanyon2008a}. This is identical to the symmetrisation needed to generate an AKLT state~\cite{Affleck1987a} (see~Fig.~1 in the main text). We measure this qutrit using the analyser outlined in Figure~\ref{setup}c. 

\begingroup
\squeezetable
\begin{table*}
 \begin{center}
 \begin{tabular}{ccccccccccccccccc}
    \hline\hline
    & \hspace{0.5em} & $\ket{\mathbf{0}}$ & $\ket{\mathbf{1}}$ & $\ket{\mathbf{2}}$
    & $\frac{\ket{\mathbf{0}}+\ket{\mathbf{1}}}{\sqrt{2}}$ 
    & $\frac{\ket{\mathbf{0}}-\ket{\mathbf{1}}}{\sqrt{2}}$ & $\frac{\ket{\mathbf{1}}+\ket{\mathbf{2}}}{\sqrt{2}}$
    & $\frac{\ket{\mathbf{1}}-\ket{\mathbf{2}}}{\sqrt{2}}$ & $\frac{\ket{\mathbf{2}}+\ket{\mathbf{0}}}{\sqrt{2}}$
    & $\frac{\ket{\mathbf{2}}-\ket{\mathbf{0}}}{\sqrt{2}}$ 
    & $\frac{\ket{\mathbf{0}}+i\ket{\mathbf{1}}}{\sqrt{2}}$ 
    & $\frac{\ket{\mathbf{0}}-i\ket{\mathbf{1}}}{\sqrt{2}}$ & $\frac{\ket{\mathbf{1}}+i\ket{\mathbf{2}}}{\sqrt{2}}$
    & $\frac{\ket{\mathbf{1}}-i\ket{\mathbf{2}}}{\sqrt{2}}$ & $\frac{\ket{\mathbf{2}}+i\ket{\mathbf{0}}}{\sqrt{2}}$
    & $\frac{\ket{\mathbf{2}}-i\ket{\mathbf{0}}}{\sqrt{2}}$ \\[0.3ex]\hline
    $\alpha_1$ 	& & 0 & 0 & $\frac{\pi}{2}$ & 0 & 0 & $\frac{\pi}{2}$ & $\frac{\pi}{2}$ & $-\frac{\pi}{4}$ & $\pi$ & 0 & 0 & $\frac{\pi}{2}$ &
		$\frac{\pi}{2}$ & $-\frac{\pi}{4}$ & $-\frac{\pi}{4}$ \\[0.3ex]
    $\chi_1$ 	& & 0 & 0 & 0 & 0 & 0 & 0 & 0 & $\frac{\pi}{2}$ & $-\frac{\pi}{4}$ & 0 & 0 & 0 & 0 & $\frac{\pi}{4}$ & $-\frac{\pi}{4}$ \\[0.3ex]
    $\alpha_2$ 	& & 0 & $\frac{\pi}{2}$ & $\frac{\pi}{2}$ & $\xi$ & $\xi$ & $\eta$ & $\eta$ & $\frac{\pi}{4}$ & 0 & $\xi$ & $\xi$ & $\eta$ & $\eta$ 
		& $\frac{\pi}{4}$ & $\frac{\pi}{4}$ \\[0.3ex]
    $\chi_2$ 	& & 0 & 0 & 0 & 0 & $\pi$ & 0 & $\pi$ & $\frac{\pi}{2}$ & $\frac{\pi}{4}$ & $\frac{\pi}{2}$ & $-\frac{\pi}{2}$ 
		& $\frac{\pi}{2}$ & $-\frac{\pi}{2}$ & $\frac{\pi}{4}$ & $-\frac{\pi}{4}$ \\[0.3ex]\hline
    $p$ 	& & $\frac{1}{2}$	& $\frac{1}{4}$	& $\frac{1}{2}$	& $\frac{1}{3}$	& $\frac{1}{3}$	& $\frac{1}{3}$	& $\frac{1}{3}$	& $\frac{1}{4}$	
		& $\frac{1}{4}$	& $\frac{1}{3}$	& $\frac{1}{3}$	& $\frac{1}{3}$	& $\frac{1}{3}$	& $\frac{1}{4}$	& $\frac{1}{4}$\\[0.3ex]\hline\hline
 \end{tabular}
 \caption{Analyser parameters for qutrit measurements used for over-complete tomography. Each qutrit measurement is implemented by splitting the biphoton probabilistically at a beam splitter and projecting the two photons ($m=1,2$) on $\ket{\psi_m} = \cos\alpha_m \ket{\SH}+e^{i\chi_m}\sin\alpha_m\ket{\SV}$. For brevity, we use the definitions $\xi = \mathrm{arccos}(\frac{1}{\sqrt{3}})$ and $\eta = \mathrm{arccos}(\sqrt{\frac{2}{3}})$, and $p$ denotes the success probability for a given qutrit projection. These success probabilities are taken into account by our tomographic reconstruction of the density matrix.\label{Q3TomoSettings}}
 \end{center}
\end{table*}
\endgroup

The analyser works by probabilistically splitting up the biphoton at a BS, and by performing a qubit projective measurement on each of the output modes $1$ and $2$ of the BS. We project on a given qutrit state by projecting the two photons on corresponding qubit states. In particular, the photon in mode $m=1,2$ of the qutrit analyser is projected on state $\ket{\psi_m} = \cos\alpha_m \ket{\SH}+e^{i\chi_m}\sin\alpha_m\ket{\SV}$. A successful projective measurement of a qutrit is heralded by a coincidence event between detectors $D_1$ and $D_2$. To calculate which qutrit state such a coincidence event signals we can propagate our two-qubit state $\ket{\psi_1}\otimes\ket{\psi_2}$ back through the beam splitter. All contributions in the unused of the two input ports of the BS can be neglected. The (unnormalised) two-photon state then becomes:

\begin{widetext}
\begin{equation}
 \frac{1}{2} \left[\cos\alpha_1 \cos\alpha_2 a^\dagger_\SH a^\dagger_\SH + \left(\mathrm{e}^{i \chi_1} \cos\alpha_2 \sin\alpha_1+\mathrm{e}^{i \chi_2} \cos\alpha_1 \sin\alpha_2\right) a^\dagger_\SH a^\dagger_\SV +\mathrm{e}^{i (\chi_1+\chi_2)}\sin\alpha_1 \sin\alpha_2 a^\dagger_\SV a^\dagger_\SV\right]\vac.
\end{equation}
In our qutrit basis we can write this as: 
\begin{equation}
 \frac{1}{\sqrt{2}} \cos\alpha_1 \cos\alpha_2 \ket{\mathbf{0}} + \frac{1}{2} \left(\mathrm{e}^{i \chi_1} \cos\alpha_2 \sin\alpha_1+\mathrm{e}^{i \chi_2} \cos\alpha_1 \sin\alpha_2\right) \ket{\mathbf{1}} + \frac{1}{\sqrt{2}} \mathrm{e}^{i (\chi_1+\chi_2)}\sin\alpha_1 \sin\alpha_2 \ket{\mathbf{2}}.
\end{equation} 
\end{widetext}

In general, the success probability of this projective qutrit measurement depends on the qutrit state. For example, to project on the biphoton state $a^\dagger_\SH a^\dagger_\SH \vac$, we set both analysers to $\ket{\SH}$. Given that the biphoton is in the correct state, a coincidence will occur with probability $\frac{1}{2}$ because of the probabilistic splitting of the photons at the BS. As a second example, in order to project on the biphoton state $a^\dagger_\SH a^\dagger_\SV \vac$, we choose $\ket{\psi_1} = \ket{\SH}$ and $\ket{\psi_2} = \ket{\SV}$. If the biphoton is in the right state, the success probability will only be $\frac{1}{4}$ because the photons can be split in four possible ways, and only one leads to a coincidence event.

Table~\ref{Q3RotSettings} lists the parameters $\alpha_m$ and $\chi_m$ we have to choose to project on a qutrit measurement for a rotation by an angle $\theta$ around axis $\hat{x}$, $\hat{y}$ or $\hat{z}$. The probability for each of these qutrit measurements to work is $\frac{1}{4}$. By taking into account the $\ket{\SV}$ outcomes of the polarising beam splitters (PBSs) in the qutrit analyser, this probability could be increased to $\frac{1}{2}$.
In order to analyse the qubit-qutrit-qubit state generated in our setup we performed a tomographically over-complete set of measurements and performed tomography using a maximum likelihood technique~\cite{James2001a,Doherty2009a}. The set of measurements performed for the two qubits was $\ket{\SH}$, $\ket{\SV}$, $\ket{\SD}$, $\ket{\SA}$, $\ket{\SR}$, and $\ket{\SL}$. In Table~\ref{Q3TomoSettings} we list the set of qutrit measurements as well as the corresponding parameters $\alpha_m$ and $\chi_m$ for the settings of the qubit measurements in the qutrit analyser. The table also shows the probability with which each of the projections succeeds given that the biphoton is in the corresponding qutrit state.

A qubit analyser where both outputs ($\ket{\SH}$ and $\ket{\SV}$) are monitored can be operated in various ways. One is to adjust the coupling efficiencies of the two single-mode fibre couplers such that they are approximately the same. This method is, however, very susceptible to long-time drifts of the sources because these might again lead to an unbalance between the two couplers. The method we chose is the following. Let us assume we want to measure the counts for a projection on $\ket{\SH}$. Then we record the counts in both PBS outputs once with the waveplates set such that the transmitted path in the PBS corresponds to $\ket{\SH}$, and we record the counts in both PBS outputs for the same amount of time with the waveplates set such that the transmitted path corresponds to $\ket{\SV}$. We do the same for any polarisation we project on, and by adding up the respective counts we average over any imperfections in the balance of the coupling efficiencies. If we have $N$ qubit analysers for which we apply this technique, the number of combinations to measure will be $2^N$. For our tomography measurements of the AKLT state we used this technique for modes $1$ and $4$. For the single-qubit rotation measurements we used it for mode $4$, but we only monitored the transmitted mode in analyser $1$ in order to simplify the software implementation of the scan routine. 

\section{Results}
\label{app:results}

\begingroup
\squeezetable
\begin{table*}
 \begin{center}
 \begin{tabular}{cccccccccccccccccc}
    \hline\hline
    & & \hspace{0.3em} & $\ket{\mathbf{0}}$ & $\ket{\mathbf{1}}$ & $\ket{\mathbf{2}}$ 
    & $\frac{\ket{\mathbf{0}}+\ket{\mathbf{1}}}{\sqrt{2}}$ 
    & $\frac{\ket{\mathbf{0}}-\ket{\mathbf{1}}}{\sqrt{2}}$ & $\frac{\ket{\mathbf{1}}+\ket{\mathbf{2}}}{\sqrt{2}}$ 
    & $\frac{\ket{\mathbf{1}}-\ket{\mathbf{2}}}{\sqrt{2}}$ & $\frac{\ket{\mathbf{2}}+\ket{\mathbf{0}}}{\sqrt{2}}$ 
    & $\frac{\ket{\mathbf{2}}-\ket{\mathbf{0}}}{\sqrt{2}}$ 
    & $\frac{\ket{\mathbf{0}}+i\ket{\mathbf{1}}}{\sqrt{2}}$ 
    & $\frac{\ket{\mathbf{0}}-i\ket{\mathbf{1}}}{\sqrt{2}}$ & $\frac{\ket{\mathbf{1}}+i\ket{\mathbf{2}}}{\sqrt{2}}$ 
    & $\frac{\ket{\mathbf{1}}-i\ket{\mathbf{2}}}{\sqrt{2}}$ & $\frac{\ket{\mathbf{2}}+i\ket{\mathbf{0}}}{\sqrt{2}}$ 
    & $\frac{\ket{\mathbf{2}}-i\ket{\mathbf{0}}}{\sqrt{2}}$ \\[0.7ex]\hline
    \multirow{6}{*}{$\ket{\SH}$} & $\ket{\SH}$ & & $3$ & $8$ & $595$ & $7$ & $7$ & $239$ & $191$ & $158$ & $128$ & $7$ & $10$ & $181$ & $198$ & $154$ & $153$\\[0.3ex]  
	& $\ket{\SV}$ & & $27$ & $195$ & $27$ & $132$ & $107$ & $110$ & $151$ & $26$ & $22$ & $141$ & $113$ & $147$ & $110$ & $18$ & $22$\\[0.3ex]
	& $\ket{\SD}$ & & $8$ & $80$ & $327$ & $63$ & $68$ & $26$ & $296$ & $90$ & $86$ & $75$ & $57$ & $171$ & $176$ & $100$ & $93$\\[0.3ex] 
	& $\ket{\SA}$ & & $9$ & $120$ & $321$ & $82$ & $74$ & $384$ & $24$ & $88$ & $67$ & $101$ & $78$ & $212$ & $160$ & $81$ & $86$\\[0.3ex]
	& $\ket{\SR}$ & & $12$ & $90$ & $251$ & $78$ & $61$ & $171$ & $158$ & $95$ & $69$ & $80$ & $61$ & $17$ & $306$ & $109$ & $71$\\[0.3ex]
	& $\ket{\SL}$ & & $10$ & $111$ & $297$ & $72$ & $62$ & $165$ & $170$ & $62$ & $110$ & $84$ & $74$ & $366$ & $22$ & $88$ & $101$\\[0.3ex]\hline
    \multirow{6}{*}{$\ket{\SV}$} & $\ket{\SH}$ & & $17$ & $127$ & $17$ & $86$ & $78$ & $77$ & $126$ & $14$ & $12$ & $65$ & $98$ & $71$ & $80$ & $12$ & $11$\\[0.3ex]
	& $\ket{\SV}$ & & $539$ & $3$ & $0$ & $141$ & $202$ & $5$ & $4$ & $153$ & $145$ & $177$ & $154$ & $3$ & $4$ & $128$ & $141$\\[0.3ex]
	& $\ket{\SD}$ & & $256$ & $79$ & $8$ & $9$ & $277$ & $43$ & $53$ & $83$ & $77$ & $146$ & $150$ & $54$ & $44$ & $60$ & $76$\\[0.3ex]
	& $\ket{\SA}$ & & $317$ & $68$ & $4$ & $204$ & $22$ & $45$ & $52$ & $102$ & $66$ & $166$ & $152$ & $34$ & $53$ & $93$ & $46$\\[0.3ex]
	& $\ket{\SR}$ & & $329$ & $61$ & $5$ & $140$ & $129$ & $34$ & $57$ & $62$ & $87$ & $23$ & $218$ & $46$ & $48$ & $90$ & $89$\\[0.3ex]
	& $\ket{\SL}$ & & $299$ & $71$ & $8$ & $113$ & $122$ & $45$ & $52$ & $82$ & $61$ & $239$ & $19$ & $62$ & $41$ & $84$ & $59$\\[0.3ex]\hline
    \multirow{6}{*}{$\ket{\SD}$} & $\ket{\SH}$ & & $5$ & $68$ & $328$ & $46$ & $40$ & $30$ & $238$ & $58$ & $85$ & $31$ & $58$ & $113$ & $160$ & $58$ & $85$\\[0.3ex]
	& $\ket{\SV}$ & & $283$ & $108$ & $11$ & $13$ & $319$ & $55$ & $96$ & $80$ & $43$ & $156$ & $160$ & $70$ & $67$ & $70$ & $61$\\[0.3ex]
	& $\ket{\SD}$ & & $127$ & $173$ & $193$ & $18$ & $231$ & $5$ & $292$ & $117$ & $21$ & $138$ & $162$ & $144$ & $155$ & $69$ & $120$\\[0.3ex]
	& $\ket{\SA}$ & & $141$ & $34$ & $168$ & $34$ & $85$ & $84$ & $42$ & $6$ & $140$ & $79$ & $52$ & $52$ & $56$ & $71$ & $83$\\[0.3ex]
	& $\ket{\SR}$ & & $126$ & $85$ & $139$ & $23$ & $171$ & $51$ & $156$ & $74$ & $84$ & $52$ & $184$ & $21$ & $192$ & $14$ & $151$\\[0.3ex]
	& $\ket{\SL}$ & & $122$ & $88$ & $205$ & $27$ & $180$ & $41$ & $162$ & $75$ & $84$ & $160$ & $49$ & $166$ & $35$ & $149$ & $20$\\[0.3ex]\hline
    \multirow{6}{*}{$\ket{\SA}$} & $\ket{\SH}$ & & $8$ & $64$ & $320$ & $80$ & $54$ & $282$ & $23$ & $67$ & $66$ & $52$ & $68$ & $149$ & $151$ & $66$ & $79$\\[0.3ex]
	& $\ket{\SV}$ & & $315$ & $87$ & $15$ & $303$ & $16$ & $67$ & $43$ & $61$ & $99$ & $198$ & $147$ & $77$ & $57$ & $80$ & $101$\\[0.3ex]
	& $\ket{\SD}$ & & $154$ & $18$ & $173$ & $67$ & $50$ & $54$ & $56$ & $14$ & $136$ & $67$ & $49$ & $68$ & $58$ & $66$ & $95$\\[0.3ex]
	& $\ket{\SA}$ & & $166$ & $174$ & $169$ & $273$ & $8$ & $286$ & $15$ & $167$ & $14$ & $181$ & $159$ & $139$ & $130$ & $75$ & $89$\\[0.3ex]
	& $\ket{\SR}$ & & $189$ & $87$ & $148$ & $166$ & $23$ & $171$ & $26$ & $80$ & $93$ & $56$ & $159$ & $30$ & $164$ & $157$ & $11$\\[0.3ex]
	& $\ket{\SL}$ & & $154$ & $89$ & $167$ & $152$ & $26$ & $179$ & $32$ & $79$ & $92$ & $196$ & $39$ & $204$ & $45$ & $20$ & $125$\\[0.3ex]\hline
    \multirow{6}{*}{$\ket{\SR}$} & $\ket{\SH}$ & & $10$ & $58$ & $270$ & $41$ & $42$ & $167$ & $139$ & $77$ & $84$ & $39$ & $45$ & $34$ & $300$ & $96$ & $80$\\[0.3ex]
	& $\ket{\SV}$ & & $298$ & $85$ & $5$ & $161$ & $119$ & $59$ & $71$ & $98$ & $68$ & $19$ & $239$ & $76$ & $65$ & $73$ & $60$\\[0.3ex]
	& $\ket{\SD}$ & & $116$ & $68$ & $169$ & $32$ & $140$ & $26$ & $181$ & $67$ & $99$ & $12$ & $153$ & $52$ & $201$ & $9$ & $127$\\[0.3ex]
	& $\ket{\SA}$ & & $159$ & $93$ & $140$ & $161$ & $57$ & $182$ & $34$ & $94$ & $71$ & $24$ & $192$ & $55$ & $168$ & $182$ & $8$\\[0.3ex]
	& $\ket{\SR}$ & & $135$ & $126$ & $183$ & $172$ & $128$ & $156$ & $149$ & $12$ & $137$ & $8$ & $255$ & $15$ & $296$ & $87$ & $89$\\[0.3ex]
	& $\ket{\SL}$ & & $116$ & $22$ & $192$ & $57$ & $46$ & $93$ & $68$ & $162$ & $18$ & $35$ & $66$ & $96$ & $62$ & $57$ & $113$\\[0.3ex]\hline
    \multirow{6}{*}{$\ket{\SL}$} & $\ket{\SH}$ & & $9$ & $63$ & $229$ & $60$ & $37$ & $142$ & $134$ & $97$ & $66$ & $56$ & $53$ & $267$ & $24$ & $47$ & $71$\\[0.3ex]
	& $\ket{\SV}$ & & $340$ & $73$ & $6$ & $165$ & $183$ & $58$ & $67$ & $84$ & $86$ & $329$ & $20$ & $86$ & $45$ & $99$ & $63$\\[0.3ex]
	& $\ket{\SD}$ & & $144$ & $85$ & $167$ & $36$ & $194$ & $36$ & $214$ & $100$ & $72$ & $182$ & $46$ & $191$ & $37$ & $143$ & $13$\\[0.3ex]
	& $\ket{\SA}$ & & $195$ & $101$ & $128$ & $167$ & $49$ & $139$ & $30$ & $97$ & $80$ & $221$ & $23$ & $172$ & $37$ & $20$ & $148$\\[0.3ex]
	& $\ket{\SR}$ & & $180$ & $13$ & $142$ & $61$ & $75$ & $57$ & $71$ & $166$ & $8$ & $100$ & $43$ & $45$ & $62$ & $68$ & $93$\\[0.3ex]
	& $\ket{\SL}$ & & $133$ & $152$ & $172$ & $149$ & $124$ & $150$ & $163$ & $10$ & $131$ & $271$ & $14$ & $306$ & $15$ & $82$ & $82$\\[0.3ex]\hline\hline
 \end{tabular}
 \caption{Table of tomography results. We performed a tomographically-overcomplete set of measurements on our qubit-qutrit-qubit state. The states in the first column indicate the measurement settings for the first qubit, the state in the second column indicate the settings for the other qubit, and the states in the top row denote the qutrit measurement settings. All counts given are raw four-fold coincidence events integrated over $480$s. We have \textit{not} corrected these counts to take into account the probabilities of success for the various qutrit measurements involved. In order to make this correction to get the ``real'' count rates, which one can use for the tomographic reconstruction, one has to divide the raw counts by the probability of success for the qutrit measurement as given in Table~\ref{Q3TomoSettings}.\label{tomocounts}}
 \end{center}
\end{table*}
\endgroup

\begin{table*}
 \begin{center}
  \begin{tabular}{ccccccccc}
    \hline\hline
   \multirow{2}{*}{outcome} & \multirow{2}{*}{$\theta$} & & \multicolumn{6}{c}{logical input state} \\[0.3ex]\cline{4-9}
      &  & & $\ket{\SH}$ & $\ket{\SV}$ & $\ket{\mathfrak{h}^+}$ & $\ket{\mathfrak{h}^-}$ & $\ket{\mathfrak{m}^+}$ & $\ket{\mathfrak{m}^-}$ \\[0.3ex]\hline
\multirow{10}{*}{plus} & $0$ & & $0.84\pm 0.03$ & $0.93\pm 0.02$ & $0.97\pm 0.03$ & $0.98\pm 0.02$ & $0.90\pm 0.03$ & $0.94\pm 0.02$\\[0.3ex]
 & $\pi/8$ & & $0.97\pm 0.03$ & $0.86\pm 0.05$ & $0.96\pm 0.02$ & $0.91\pm 0.06$ & $0.94\pm 0.02$ & $0.95\pm 0.02$\\[0.3ex]
 & $\pi/4$ & & $0.85\pm 0.05$ & $0.93\pm 0.03$ & $0.93\pm 0.04$ & $0.88\pm 0.05$ & $0.98\pm 0.02$ & $0.93\pm 0.02$\\[0.3ex]
 & $3\pi/8$ & & $0.90\pm 0.03$ & $0.93\pm 0.03$ & $0.93\pm 0.03$ & $0.92\pm 0.04$ & $0.94\pm 0.03$ & $0.86\pm 0.03$\\[0.3ex]
 & $\pi/2$ & & $0.91\pm 0.03$ & $0.82\pm 0.06$ & $0.91\pm 0.03$ & $0.91\pm 0.05$ & $0.90\pm 0.03$ & $0.77\pm 0.03$\\[0.3ex]
 & $3\pi/4$ & & $0.88\pm 0.05$ & $0.91\pm 0.05$ & $0.92\pm 0.05$ & $0.90\pm 0.06$ & $0.86\pm 0.03$ & $0.94\pm 0.03$\\[0.3ex]
 & $\pi$ & & $0.97\pm 0.01$ & $0.96\pm 0.01$ & $0.91\pm 0.03$ & $0.88\pm 0.03$ & $0.87\pm 0.04$ & $0.86\pm 0.04$\\[0.3ex]
 & $5\pi/4$ & & $0.92\pm 0.03$ & $0.93\pm 0.03$ & $0.81\pm 0.04$ & $0.94\pm 0.03$ & $0.94\pm 0.03$ & $0.95\pm 0.02$\\[0.3ex]
 & $3\pi/2$ & & $0.90\pm 0.02$ & $0.95\pm 0.01$ & $0.90\pm 0.03$ & $0.98\pm 0.02$ & $0.85\pm 0.04$ & $0.95\pm 0.03$\\[0.3ex]
 & $7\pi/4$ & & $0.90\pm 0.04$ & $0.94\pm 0.03$ & $0.89\pm 0.04$ & $0.97\pm 0.02$ & $0.94\pm 0.03$ & $0.89\pm 0.03$\\[0.3ex]
\hline
\multirow{10}{*}{minus} & $0$ & & $0.96\pm 0.02$ & $0.92\pm 0.03$ & $0.87\pm 0.05$ & $0.84\pm 0.05$ & $0.94\pm 0.03$ & $0.88\pm 0.03$\\[0.3ex]
 & $\pi/8$ & & $0.94\pm 0.03$ & $0.98\pm 0.03$ & $0.83\pm 0.04$ & $0.84\pm 0.05$ & $0.90\pm 0.03$ & $0.90\pm 0.03$\\[0.3ex]
 & $\pi/4$ & & $0.93\pm 0.03$ & $0.99\pm 0.03$ & $0.96\pm 0.02$ & $0.91\pm 0.05$ & $0.95\pm 0.03$ & $0.97\pm 0.01$\\[0.3ex]
 & $3\pi/8$ & & $0.93\pm 0.04$ & $0.94\pm 0.03$ & $0.98\pm 0.02$ & $0.91\pm 0.04$ & $0.94\pm 0.03$ & $0.88\pm 0.03$\\[0.3ex]
 & $\pi/2$ & & $0.89\pm 0.03$ & $0.95\pm 0.02$ & $0.88\pm 0.05$ & $0.90\pm 0.04$ & $0.92\pm 0.04$ & $0.89\pm 0.04$\\[0.3ex]
 & $3\pi/4$ & & $0.99\pm 0.01$ & $0.96\pm 0.03$ & $0.91\pm 0.04$ & $0.93\pm 0.04$ & $0.97\pm 0.02$ & $0.98\pm 0.02$\\[0.3ex]
 & $\pi$ & & $0.89\pm 0.02$ & $0.91\pm 0.02$ & $0.85\pm 0.04$ & $0.96\pm 0.02$ & $0.94\pm 0.03$ & $0.87\pm 0.03$\\[0.3ex]
 & $5\pi/4$ & & $0.90\pm 0.03$ & $0.97\pm 0.02$ & $0.86\pm 0.04$ & $0.88\pm 0.04$ & $0.95\pm 0.03$ & $0.93\pm 0.03$\\[0.3ex]
 & $3\pi/2$ & & $0.92\pm 0.02$ & $0.94\pm 0.02$ & $0.91\pm 0.03$ & $0.87\pm 0.04$ & $0.88\pm 0.04$ & $0.89\pm 0.03$\\[0.3ex]
 & $7\pi/4$ & & $0.94\pm 0.02$ & $0.98\pm 0.03$ & $0.92\pm 0.03$ & $0.83\pm 0.04$ & $0.95\pm 0.02$ & $0.92\pm 0.02$\\[0.3ex]
\hline
id & n.a. &  & $0.90\pm 0.02$ & $0.90\pm 0.02$ & $0.88\pm 0.04$ & $0.96\pm 0.02$ & $0.90\pm 0.03$ & $0.93\pm 0.03$\\[0.3ex]\hline\hline
  \end{tabular}
  \caption{Fidelities for rotations around the $\hat{x}$ axis. This table lists the fidelities for various input states of the computation compared with the theoretical expectation given an ideal AKLT state. These fidelities are given for a number of angles of rotation around the $\hat{x}$ axis. We list them separately for the three distinct outcomes `plus', `minus' and `id' of the qutrit measurement. For each input state we measured the `id' outcome only once because it is independent of the rotation angle. The value and error for each fidelity are the mean and standard deviation from Monte Carlo simulations based on a Poissonian distribution around the counts measured.\label{rotationfidelitiesX}} 
 \end{center}

\end{table*}

\begin{table*}
 \begin{center}
  \begin{tabular}{ccccccccc}
    \hline\hline
    \multirow{2}{*}{outcome} & \multirow{2}{*}{$\theta$} & & \multicolumn{6}{c}{logical input state} \\[0.3ex]\cline{4-9}
      & & & $\ket{\SH}$ & $\ket{\SV}$ & $\ket{\mathfrak{h}^+}$ & $\ket{\mathfrak{h}^-}$ & $\ket{\mathfrak{m}^+}$ & $\ket{\mathfrak{m}^-}$ \\[0.3ex]\hline
\multirow{10}{*}{plus} & $0$ & & $0.95\pm 0.01$ & $0.91\pm 0.02$ & $0.92\pm 0.03$ & $0.88\pm 0.04$ & $0.94\pm 0.03$ & $0.95\pm 0.02$\\[0.3ex]
 & $\pi/8$ & & $0.93\pm 0.02$ & $0.94\pm 0.02$ & $0.92\pm 0.03$ & $0.91\pm 0.03$ & $0.94\pm 0.03$ & $0.85\pm 0.03$\\[0.3ex]
 & $\pi/4$ & & $0.94\pm 0.03$ & $0.90\pm 0.04$ & $0.91\pm 0.02$ & $0.87\pm 0.02$ & $0.93\pm 0.03$ & $0.98\pm 0.02$\\[0.3ex]
 & $3\pi/8$ & & $0.92\pm 0.03$ & $0.95\pm 0.02$ & $0.94\pm 0.02$ & $0.92\pm 0.03$ & $0.93\pm 0.02$ & $0.86\pm 0.03$\\[0.3ex]
 & $\pi/2$ & & $0.89\pm 0.02$ & $0.91\pm 0.02$ & $0.85\pm 0.03$ & $0.88\pm 0.03$ & $0.84\pm 0.04$ & $0.87\pm 0.03$\\[0.3ex]
 & $3\pi/4$ & & $0.93\pm 0.03$ & $0.86\pm 0.04$ & $0.84\pm 0.03$ & $0.86\pm 0.03$ & $0.89\pm 0.03$ & $0.89\pm 0.02$\\[0.3ex]
 & $\pi$ & & $0.83\pm 0.03$ & $0.85\pm 0.03$ & $0.87\pm 0.04$ & $0.94\pm 0.03$ & $0.90\pm 0.03$ & $0.94\pm 0.03$\\[0.3ex]
 & $5\pi/4$ & & $0.80\pm 0.04$ & $0.88\pm 0.03$ & $0.90\pm 0.02$ & $0.95\pm 0.02$ & $0.97\pm 0.02$ & $0.95\pm 0.02$\\[0.3ex]
 & $3\pi/2$ & & $0.87\pm 0.03$ & $0.88\pm 0.02$ & $0.87\pm 0.04$ & $0.93\pm 0.03$ & $0.95\pm 0.03$ & $0.90\pm 0.03$\\[0.3ex]
 & $7\pi/4$ & & $0.92\pm 0.03$ & $0.90\pm 0.03$ & $0.91\pm 0.02$ & $0.93\pm 0.02$ & $0.93\pm 0.03$ & $0.99\pm 0.01$\\[0.3ex]
\hline
\multirow{10}{*}{minus} & $0$ & & $0.90\pm 0.02$ & $0.86\pm 0.02$ & $0.89\pm 0.04$ & $0.86\pm 0.03$ & $0.94\pm 0.03$ & $0.91\pm 0.03$\\[0.3ex]
 & $\pi/8$ & & $0.91\pm 0.03$ & $0.89\pm 0.03$ & $0.90\pm 0.03$ & $0.92\pm 0.02$ & $0.98\pm 0.02$ & $0.98\pm 0.02$\\[0.3ex]
 & $\pi/4$ & & $0.93\pm 0.03$ & $0.90\pm 0.03$ & $0.92\pm 0.02$ & $0.93\pm 0.02$ & $0.94\pm 0.02$ & $0.90\pm 0.03$\\[0.3ex]
 & $3\pi/8$ & & $0.91\pm 0.03$ & $0.89\pm 0.02$ & $0.92\pm 0.03$ & $0.89\pm 0.03$ & $0.91\pm 0.03$ & $0.95\pm 0.02$\\[0.3ex]
 & $\pi/2$ & & $0.91\pm 0.02$ & $0.91\pm 0.02$ & $0.96\pm 0.03$ & $0.85\pm 0.03$ & $0.91\pm 0.03$ & $0.96\pm 0.02$\\[0.3ex]
 & $3\pi/4$ & & $0.90\pm 0.04$ & $0.96\pm 0.03$ & $0.93\pm 0.02$ & $0.89\pm 0.03$ & $0.95\pm 0.02$ & $0.93\pm 0.02$\\[0.3ex]
 & $\pi$ & & $0.91\pm 0.02$ & $0.94\pm 0.02$ & $0.89\pm 0.03$ & $0.92\pm 0.04$ & $0.95\pm 0.02$ & $0.95\pm 0.02$\\[0.3ex]
 & $5\pi/4$ & & $0.95\pm 0.02$ & $0.99\pm 0.01$ & $0.89\pm 0.02$ & $0.84\pm 0.03$ & $0.91\pm 0.03$ & $0.96\pm 0.02$\\[0.3ex]
 & $3\pi/2$ & & $0.92\pm 0.02$ & $0.86\pm 0.02$ & $0.84\pm 0.04$ & $0.82\pm 0.03$ & $0.90\pm 0.03$ & $0.91\pm 0.03$\\[0.3ex]
 & $7\pi/4$ & & $0.83\pm 0.04$ & $0.90\pm 0.03$ & $0.92\pm 0.02$ & $0.91\pm 0.02$ & $0.86\pm 0.03$ & $0.95\pm 0.02$\\[0.3ex]
\hline
id & n.a. &  & $0.92\pm 0.02$ & $0.88\pm 0.02$ & $0.89\pm 0.04$ & $0.93\pm 0.03$ & $0.95\pm 0.03$ & $0.93\pm 0.03$\\[0.3ex]\hline\hline
  \end{tabular}
  \caption{Fidelities for rotations around the $\hat{y}$ axis.\label{rotationfidelitiesY}} 
 \end{center}

\end{table*}

\begin{table*}
 \begin{center}
  \begin{tabular}{ccccccccccc}
    \hline\hline
    \multirow{2}{*}{outcome} & \multirow{2}{*}{$\theta$} & & \multicolumn{6}{c}{logical input state} \\[0.3ex]\cline{4-11}
      & & & $\ket{\SH}$ & $\ket{\SV}$ & $\ket{\mathfrak{h}^+}$ & $\ket{\mathfrak{h}^-}$ & $\ket{\mathfrak{m}^+}$ & $\ket{\mathfrak{m}^-}$ & $\ket{\SD}$ & $\ket{\SA}$ \\[0.3ex]\hline
\multirow{10}{*}{plus} & $0$ & & $0.87\pm 0.03$ & $0.94\pm 0.02$ & $0.89\pm 0.04$ & $0.89\pm 0.04$ & $0.93\pm 0.03$ & $0.91\pm 0.03$ & $0.92\pm 0.03$ & $0.90\pm 0.03$\\[0.3ex]
 & $\pi/8$ & & $0.92\pm 0.02$ & $0.91\pm 0.02$ & $0.86\pm 0.04$ & $0.92\pm 0.03$ & $0.92\pm 0.03$ & $0.93\pm 0.03$ & $0.97\pm 0.02$ & $0.98\pm 0.01$\\[0.3ex]
 & $\pi/4$ & & $0.85\pm 0.03$ & $0.85\pm 0.03$ & $0.91\pm 0.04$ & $0.89\pm 0.04$ & $0.94\pm 0.03$ & $0.94\pm 0.03$ & $0.97\pm 0.03$ & $0.89\pm 0.04$\\[0.3ex]
 & $3\pi/8$ & & $0.85\pm 0.03$ & $0.95\pm 0.01$ & $0.88\pm 0.04$ & $0.89\pm 0.03$ & $0.86\pm 0.04$ & $0.91\pm 0.03$ & $0.92\pm 0.03$ & $0.88\pm 0.04$\\[0.3ex]
 & $\pi/2$ & & $0.90\pm 0.03$ & $0.92\pm 0.02$ & $0.79\pm 0.06$ & $0.98\pm 0.01$ & $0.85\pm 0.04$ & $0.86\pm 0.03$ & $0.88\pm 0.05$ & $0.89\pm 0.04$\\[0.3ex]
 & $3\pi/4$ & & $0.86\pm 0.04$ & $0.93\pm 0.02$ & $0.95\pm 0.04$ & $0.93\pm 0.04$ & $0.83\pm 0.04$ & $0.88\pm 0.04$ & $0.92\pm 0.04$ & $0.87\pm 0.06$\\[0.3ex]
 & $\pi$ & & $0.92\pm 0.03$ & $0.87\pm 0.03$ & $0.79\pm 0.05$ & $0.84\pm 0.05$ & $0.91\pm 0.04$ & $0.90\pm 0.04$ & $0.95\pm 0.03$ & $0.96\pm 0.02$\\[0.3ex]
 & $5\pi/4$ & & $0.91\pm 0.03$ & $0.87\pm 0.04$ & $0.98\pm 0.01$ & $0.91\pm 0.05$ & $0.89\pm 0.03$ & $0.99\pm 0.01$ & $0.93\pm 0.04$ & $0.96\pm 0.04$\\[0.3ex]
 & $3\pi/2$ & & $0.90\pm 0.04$ & $0.95\pm 0.02$ & $0.92\pm 0.05$ & $0.90\pm 0.05$ & $0.94\pm 0.03$ & $0.88\pm 0.04$ & $0.91\pm 0.03$ & $0.93\pm 0.03$\\[0.3ex]
 & $7\pi/4$ & & $0.94\pm 0.03$ & $0.99\pm 0.01$ & $0.91\pm 0.06$ & $0.99\pm 0.01$ & $0.93\pm 0.03$ & $0.96\pm 0.03$ & $0.83\pm 0.06$ & $0.96\pm 0.04$\\[0.3ex]
\hline
\multirow{10}{*}{minus} & $0$ & & $0.87\pm 0.03$ & $0.95\pm 0.02$ & $0.88\pm 0.04$ & $0.88\pm 0.03$ & $0.90\pm 0.03$ & $0.98\pm 0.02$ & $0.90\pm 0.03$ & $0.92\pm 0.03$\\[0.3ex]
 & $\pi/8$ & & $0.89\pm 0.02$ & $0.90\pm 0.03$ & $0.93\pm 0.03$ & $0.96\pm 0.03$ & $0.94\pm 0.03$ & $0.94\pm 0.03$ & $0.92\pm 0.03$ & $0.88\pm 0.04$\\[0.3ex]
 & $\pi/4$ & & $0.95\pm 0.01$ & $0.90\pm 0.02$ & $0.92\pm 0.04$ & $0.89\pm 0.04$ & $0.96\pm 0.02$ & $0.97\pm 0.03$ & $0.89\pm 0.05$ & $0.94\pm 0.04$\\[0.3ex]
 & $3\pi/8$ & & $0.91\pm 0.02$ & $0.93\pm 0.02$ & $0.96\pm 0.03$ & $0.92\pm 0.04$ & $0.90\pm 0.03$ & $0.94\pm 0.03$ & $0.90\pm 0.04$ & $0.94\pm 0.04$\\[0.3ex]
 & $\pi/2$ & & $0.90\pm 0.03$ & $0.89\pm 0.03$ & $0.96\pm 0.03$ & $0.85\pm 0.06$ & $0.90\pm 0.04$ & $0.91\pm 0.03$ & $0.91\pm 0.03$ & $0.94\pm 0.03$\\[0.3ex]
 & $3\pi/4$ & & $0.93\pm 0.02$ & $0.94\pm 0.02$ & $0.99\pm 0.02$ & $0.93\pm 0.04$ & $0.91\pm 0.04$ & $0.98\pm 0.01$ & $0.97\pm 0.03$ & $0.99\pm 0.01$\\[0.3ex]
 & $\pi$ & & $0.86\pm 0.04$ & $0.87\pm 0.03$ & $0.93\pm 0.04$ & $0.92\pm 0.04$ & $0.90\pm 0.04$ & $0.90\pm 0.04$ & $0.92\pm 0.03$ & $0.90\pm 0.04$\\[0.3ex]
 & $5\pi/4$ & & $0.88\pm 0.05$ & $0.99\pm 0.01$ & $0.94\pm 0.05$ & $0.89\pm 0.05$ & $0.89\pm 0.04$ & $0.94\pm 0.03$ & $0.90\pm 0.05$ & $0.80\pm 0.06$\\[0.3ex]
 & $3\pi/2$ & & $0.96\pm 0.03$ & $0.91\pm 0.04$ & $0.83\pm 0.09$ & $0.98\pm 0.02$ & $0.79\pm 0.04$ & $0.86\pm 0.03$ & $0.97\pm 0.02$ & $0.96\pm 0.02$\\[0.3ex]
 & $7\pi/4$ & & $0.98\pm 0.01$ & $0.99\pm 0.01$ & $0.93\pm 0.05$ & $0.87\pm 0.06$ & $0.92\pm 0.03$ & $0.83\pm 0.04$ & $0.88\pm 0.05$ & $0.94\pm 0.04$\\[0.3ex]
\hline
id & n.a. &  & $0.96\pm 0.01$ & $0.97\pm 0.01$ & $0.96\pm 0.03$ & $0.96\pm 0.02$ & $0.93\pm 0.03$ & $0.98\pm 0.02$ & $0.90\pm 0.03$ & $0.90\pm 0.03$\\[0.3ex]\hline\hline
  \end{tabular}
  \caption{Fidelities for rotations around the $\hat{z}$ axis.\label{rotationfidelitiesZ}} 
 \end{center}

\end{table*}

All measured counts correspond to four-fold coincidence detection events between one detector in each of the two qubit analysers and the two detectors $D_1$ and $D_2$ in the qutrit analyser (see Fig.~\ref{setup}). Table~\ref{tomocounts} lists the four-fold coincidence counts measured for all tomographical settings. For each setting of the analyser waveplates we integrated over $60$s. Because we monitored both outputs in each of the two qubit analysers and applied the technique described above in order to average over any unbalance between the two analyser outputs, we have to measure $4$ combinations of settings per projective measurement. This results in $240$s overall measurement time per projective measurement. In order to reduce the effect of slow drifts in the setup, we performed the full set of measurements twice, in each case randomly ordering the settings, resulting in a measurement time of $480$s per setting in Table~\ref{tomocounts}.

In the main text we gave the measurement results for the rotation of a logical input state $\ket{\SH}$ around all three coordinate axes. For completeness, we have done the same measurements for a set of logical input states. It is important to note, that in our setup it is possible to rotate an arbitrary input state around any of the three coordinate axes because we project the first qubit on a given state rather than on a measurement basis. As we mentioned in the main text we prepare a logical input state $\ket{\psi}$ by projecting on the qubit state orthogonal to it, i.e.~$\ket{\psi^\bot}$. Because we post select on four-fold coincidence events, we automatically disregard those cases where the outcome of our measurement would be $\ket{\psi}$. If one takes into account both outcomes of the projective measurement, each of them will occur randomly with probability $\frac{1}{2}$, and if we get the outcome $\ket{\psi}$, the logical input state will be the state orthogonal to what we want to prepare. To correct for this error for any arbitrary state is impossible because that would require a universal-NOT operation, which is non-unitary~\cite{Bechmann1999a}. In practice, one chooses input states such that errors in the preparation can be corrected via Pauli operations, i.e.~any of the states along the coordinate axes of the Bloch sphere.

For each input state, rotation angle, qutrit outcome and rotation angle we performed a tomographically over-complete set of measurements ($\ket{\SH}$, $\ket{\SV}$, $\ket{\SD}$, $\ket{\SA}$, $\ket{\SR}$, and $\ket{\SL}$) on the qubit carrying the result of our single-qubit logic gate. In Tables \ref{rotationfidelitiesX}, \ref{rotationfidelitiesY}, and \ref{rotationfidelitiesZ} we list the fidelities of the reconstructed density matrices for all rotations of various input states compared with what we would expect given a perfect AKLT state. Table~\ref{rotationfidelitiesX} shows these fidelities for rotations $R_x(\theta)$ around the $\hat{x}$ axis by an angle $\theta$. For each of the $6$ different logical input states and each angle we observed the `plus' as well as the `minus' outcome of the qutrit measurement. Because the `id' outcome is independent of the rotation angle, we only measured it once for every rotation axis. For each measurement we performed a Monte Carlo simulation on the measured counts, which are assumed to be the mean of a Poissonian count distribution, to reconstruct a set of $400$ density matrices. These we used to calculate the means and standard deviations for the fidelities given in this table. Corresponding results for rotations $R_y(\theta)$ are given in Table~\ref{rotationfidelitiesY}. For rotations $R_z(\theta)$ around the $\hat{z}$ axis (see Table~\ref{rotationfidelitiesZ}) we performed measurements for two additional logical input states, $\ket{\mathbf{\pm}}$.

\bibliography{aklt_references}

\end{document}